\documentclass[a4paper]{article}
\usepackage{amsmath}   
\usepackage{amssymb}   
\usepackage{graphicx}  
\usepackage{psfrag}    
\title{Theta-vacuum and large $N$ limit \\
  in $\mathbb{C}\mathrm{P}^{N-1}$ $\sigma$ models}
\author{M.~Aguado$^1$\footnote{\texttt{miguel.aguado@mpq.mpg.de}},
  M.~Asorey$^2$\footnote{\texttt{asorey@saturno.unizar.es}} \\
  $^1$Max-Planck-Institut f\"ur Quantenoptik \\
  Hans-Kopfermann-Str.~1, D-85748 Garching (Germany) \\
  $^2$Departamento de F\'{\i}sica Te\'orica, Facultad de Ciencias \\
  Universidad de Zaragoza,
  E-50009 Zaragoza (Spain) }
\date{}
\begin{document}
%
%
\def\tilde{\widetilde}
\def\Residuo{\mathop{\rm Res}\limits}
\def\Res#1{\Residuo_{{#1}}\;}
\def\cpn{{$ \mathbb{C}\mathrm{P}^{N-1} $}}
\def\cpuno{{$ \mathbb{C}\mathrm{P}^1 $}}
\def\cpdos{{$ \mathbb{C}\mathrm{P}^2 $}}
%
%
%
\maketitle
%
%
%
\begin{abstract}
  The $\theta$ dependence of the vacuum energy density in \cpn{}
  models is re-analysed in the semiclassical approach, the $1/N$
  expansion and arguments based on the nodal structure of vacuum
  wavefunctionals.  The $1/N$ expansion is shown not to be in
  contradiction with instanton physics at finite (spacetime) volume
  $V$.  The interplay of large volume $V$ and large $N$ parameter
  gives rise to two regimes with different $\theta$ dependence, one
  behaving as a dilute instanton gas and the other dominated by the
  traditional large $N$ picture, where instantons reappear as
  resonances of the one-loop effective action, even in the absence of
  regular instantonic solutions.  The realms of the two regimes are
  given in terms of the mass gap $m$ by $m^2 V\ll N$ and $m^2 V\gg N$,
  respectively.  The small volume regime $m^2 V\ll N$ is relevant for
  physical effects associated to the physics of the boundary, like the
  leading r\^ole of edge states in the quantum Hall effect, which,
  however, do not play any r\^ole in the thermodynamic limit at large
  $N$.  Depending on the order in which the limits $N \rightarrow
  \infty$ and $V \rightarrow \infty$ are taken, two different theories
  are obtained; this is the hallmark of a phase transition at $1/N =
  0$.
\end{abstract}
%
%
\section{\label{sec:intro}%
  Introduction}

A lack of nonperturbative analytical methods haunts the study of the
infrared behaviour of confining field theories such as QCD. The main
tools used for this purpose rely on approximations (e.g.,
semiclassical, large number of colours), and rigorous results are
attainable in few corners of parameter space.  The bordering region
between topology and field physics is especially troubling, since
different methods arise sometimes from apparently incompatible
hypotheses and physical pictures.

Two-dimensional \cpn{} sigma models \cite{Eichenherr:1978qa,
  Golo:1978de, D'Adda:1978uc} are regarded as a convenient testing
ground to prepare the assault on four-dimensional gauge theories,
because both kinds of theories share a number of important properties:
conformal invariance at the classical level, asymptotic freedom,
dynamical mass generation, confinement, existence of a topological
term $\theta$ and instantons for all values of the \textsl{number of
  colours} $N$.

A relevant problem in these theories with topological properties is
the $\theta$ dependence of the vacuum energy density, the quantity
that determines the phase structure of the theory (for a recent
review, see \cite{vicarireview}).  In particular, the fate of the
discrete parity symmetry $\mathrm{P}$ upon quantisation at the values
$\theta = 0$ and $\theta = \pi$ (the only values for which it is
classically conserved) is an issue.

Perhaps the simplest model in which the subtlety of the
$\theta$-dependence of the vacuum energy $\mathcal{E}_0$ is manifest
is the quantum rotor \cite{rotor}, i.e., the quantum mechanical
problem of the dynamics of a charged particle on the circumference
$\mathrm{S}^1$ enclosing a magnetic flux $\theta$.  In the absence of
perturbations, the vacuum energy is quadratic in $\theta$; periodicity
of the physics in $\theta \rightarrow \theta + 2 \pi$ imposes that the
ground level is twofold degenerate for $\theta = \pi$ (i.e., half a
flux quantum across the region bounded by $\mathrm{S}^1$) and there
parity is spontaneously broken.  This, as we will see, mimicks the
traditional picture of the large $N$ expansion in \cpn{} models.
However, even slight perturbations compatible with reflection symmetry
lift the degeneracy of the rotor, by a level repulsion mechanism,
making the curve $\mathcal{E}_0 (\theta)$ smooth at $\theta = \pi$.  A
convenient approximate method is that of the dilute instanton gas,
where the vacuum is understood in terms of tunnelling processes among
classical vacua.  In the dilute approximation, the vacuum energy (a
pure nonperturbative effect) is a smooth periodic function of $\theta$
proportional to $(1 - \cos \theta)$.  This corresponds to the
semiclassical approximation in \cpn{} models, where instantons play an
all-important r\^ole.

\begin{figure}
  \centering
  \includegraphics{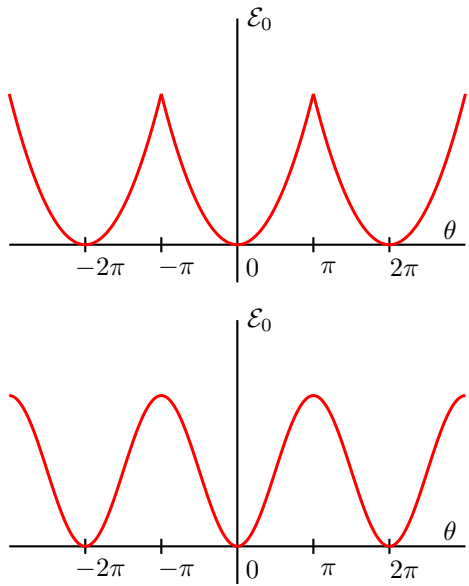}
  \caption{\label{analisisn:fig:tworegimes}%
    Structure of the vacuum energy density in the traditional large
    $N$ picture (above) and in the semiclassical picture dominated by
    instantons (below).}
\end{figure}
These two regimes have the following paradigmatic
expressions for $\mathcal{E}_0 (\theta)$, illustrated in figure
\ref{analisisn:fig:tworegimes}:
\begin{align}
\label{tworegimes:largen}
  \mathcal{E}_0 ( \theta )
&\propto
  \min \big\{ ( \theta + 2 \pi k )^2; \, k \in \mathbb{Z} \big\}
&
  \text{(large $N$),}
\\
\label{tworegimes:semiclassical}
  \mathcal{E}_0 ( \theta )
&\propto
  1 - \cos \theta 
&
  \text{(semiclassical).}
\end{align}

We now consider the situation in \cpn{} models.  Exact solutions are
known for the quantum \cpuno{} model (equivalent to the $\mathrm{O}
(3)$ model) both at $\theta = 0$ and $\theta = \pi$. In the first case
\cite{Wiegmann:1985jt}, the solution exhibits a mass gap, the spectrum
consists of an $\mathrm{SU} (2)$ triplet, and parity is conserved.
This agrees with the Haldane map \cite{Haldane:1983ru}, which
transforms this model into a chain of integer classical spins.  Vafa
and Witten \cite{Vafa:1984xg} argued that there is no first order
phase transition with spontaneous parity breaking at $\theta = 0$ for
QCD, their argument being applicable straightforwardly to all \cpn{}
models (see \cite{aa2009} for a proof of the Vafa-Witten theorem using
the topological charge as an order parameter).

The exact solution of the quantum \cpuno{} model at $\theta = \pi$
\cite{Zamolodchikov:1992zr} also conserves parity but shows no mass
gap (this result was anticipated in \cite{shankarread}).  The critical
behaviour of the model is described by an $\mathrm{SU} (2)$ WZNW model
at level $k = 1$.  This also agrees with the Haldane map, which
transforms this model into a chain of half-odd spins.  By the
Lieb-Schulz-Mattis theorem \cite{Lieb:1961fr}, the absence of mass gap
implies that P is conserved.  On the other hand, the absence of a
first order phase transition with spontaneous $\mathrm{P}$ breakdown
at $\theta = \pi$ has been argued to hold for all \cpn{} models
\cite{Asorey:1998cz}, by analyzing the nodal structure of the vacuum
in the Hamiltonian formalism \cite{Asorey:1998ciemat} in analogy with
QCD \cite{Asorey:1996pv}.

For the intermediate region $0 < \theta < \pi$, analytical techniques
are lacking, and we must rely on approximations and numerical
simulations. We will discuss two important approximations, which have
been argued to be mutually incompatible: the semiclassical method and
the $1/N$ expansion.

The semiclassical approach \cite{'tHooft:1976fv} is based, as in the
case of the rotor, on the picture of the quantum vacuum of 4$d$ gauge
theories and 2$d$ \cpn{} models built from tunnelling processes among
classical vacua.  These nonperturbative processes are dominated by
instantons and antiinstantons, (anti)selfdual solutions of the
classical Euclidean equations of motion.  A dilute gas approximation
gives a $\theta$ dependence of the vacuum energy density of the form
\begin{equation}\label{dependinstantonica}
  \mathcal{E}_0 ( \theta )
\propto
  m^2 \,
  ( 1 - \cos \theta ) ,
\end{equation}
where $m$ is the mass gap.  This dependence cannot be seen in
perturbation theory due to the nonanalytic dependence of the mass gap
on the coupling.

However, a vacuum based on a dilute gas of instantons and
antiinstantons is not satisfactory, since the statistical ensemble is
dominated by the infrared divergent contribution of arbitrarily large
instantons, whose density $n$ as a function of size $\rho$ is
\begin{equation}\label{integralinstantons}
  n ( \rho ) \mathrm{d} \rho
\propto
  ( \Lambda \rho )^N \frac{ \mathrm{d} \rho }{ \rho^3 }
\end{equation}
for the \cpn{} model, with $\Lambda$ a typical scale of the theory.  A
statistical mechanical treatment of \textsl{interacting} instanton
fluids has been developed \cite{Diakonov:1984hh, Diakonov:1986eg},
bringing about the instanton liquid picture of the QCD vacuum
\cite{Shuryak:1982pendiente}.  This may be very relevant for the
behaviour of these theories at finite temperature and high density.
We note in passing that the dilute gas approximation breaks down as
well in the \textsl{ultraviolet} for the \cpuno{} model, as pointed
out by L\"uscher \cite{luescher} building on his work with Berg
\cite{bergluescher} on the geometric definition of a topological
charge density on the lattice.  Technically, the topological
susceptibility in this model does not scale according to the
perturbative renormalisation group due to small distance fluctuations.
This is reflected in the singularity of (\ref{integralinstantons}) as
$\rho \rightarrow 0$ for $N = 2$; it may also be understood a
consequence of the slow vanishing rate of the density of Lee-Yang
zeros as $\theta\to 0$ \cite{aa2009}.  What is remarkable is that,
although the semiclassical analysis does not reveal any ultraviolet
instanton singularity in the case of \cpdos{} model, numerical
simulations suggest a similar pathology \cite{lian}.

The $1/N$ expansion \cite{'tHooft:1974hx, 'tHooft:1974jz} stands as an
alternative to the semiclassical method.  This technique is based on
the simplification of both 4$d$ SU$(N)$ gauge theories and 2$d$ \cpn{}
models when $N$ is taken to infinity keeping certain parameter
combinations constant.

The $1/N$ expansion of \cpn{} models, as developed in
\cite{D'Adda:1978uc} and \cite{Witten:1979bc}, agrees at $\theta = 0$
with the known spectrum, given by a massive particle in the adjoint
representation of $\mathrm{SU} (N)$. The mass $m_{\mathrm{T}}$ is
generated dynamically, and the particle turns out to be a composite
state of two fundamental fields, bound together by a Coulomb
potential.  For $\theta \neq 0$, this analysis predicts a quadratic
$\theta$ dependence
\begin{equation}\label{intro:ethetacuadratica}
  \mathcal{E}_0 (\theta)
=
  \frac{ 3 }{ 2 \pi } \,
  \frac{ m_{\mathrm{T}}^2 \theta^2 }{ N }
\end{equation}
of the vacuum energy density around $\theta = 0$.  This dependence can
be made to agree with the fundamental requirement that physics be
periodic in $\theta$ with period $2 \pi$ only if there is a first
order cusp at odd multiples of $\theta = \pi$, i.e., a first order
phase transition accompanied by spontaneous parity breakdown, as shown
in the upper part of figure \ref{analisisn:fig:tworegimes}.

Instanton effects, being nonperturbative, are not visible in the
perturbative expression (\ref{intro:ethetacuadratica}).  This led
Witten \cite{Witten:1979bc} to argue that the $1/N$ expansion is not
sensitive to instantons --- equivalently, that instantons play no
significant r\^ole in the quantum \cpn{} models (or in 4$d$ gauge
theories) to the extent that the $1/N$ expansion is a good
approximation thereof.  Jevicki \cite{Jevicki:1979db}, however, argued
that instantons resurface in the $1/N$ expansion as \textsl{poles} of
the integrand of the partition function $Z$
(eqn.~\ref{analisisn:funciondeparticionaalfa}), and that $Z$ can be
computed both by the saddle point method and by using a functional
Cauchy theorem summing the residues of all these poles (representing
resonances).  Then the large $N$ limit and instanton effects would not
be \textsl{a priori} incompatible with each other.

The quadratic dependence (\ref{intro:ethetacuadratica}) agrees with
the holographic picture provided by the Maldacena conjecture
\cite{Maldacena:1998re}, and moreover with lattice measurements of the
topological susceptibility of \cpn{} models (see \cite{vicarireview}
for a review).  The Witten-Veneziano formula \cite{Witten:1979vv,
  Veneziano:1979ec}, derived in this approximation, gives a
phenomenologically correct value of the $\eta'$ mass in terms of the
topological susceptibility at $\theta = 0$.  However, the appearance
of a first order cusp at $\theta = \pi$ is in contradiction with the
results arising from the nodal analysis of the vacuum
\cite{Asorey:1998cz}, and with the intuition that level repulsion
generically destroys level crossings.

In this work we show how this discrepancy stems from the fact that the
large $N$ limit and the thermodynamical limit do not commute.  The
traditional formulation of the $1/N$ expansion starts directly at
infinite spacetime volume $V = LT = \infty$.  As we shall see, a
procedure in which the thermodynamic limit is taken after the $N
\rightarrow \infty$ limit provides results compatible both with
instanton physics and with the rigorous results at $\theta = \pi$, and
different from the reverse order of limits.  A finite volume analysis
is in order.  This agrees with Schwab's \cite{Schwab1, Schwab2} and
M\"unster's \cite{Munster:1982wg, Munster:1983sd} approach in the case
of the sphere; we moreover outline the application of Jevicki's
residue method.  We find the case of the torus much more tractable and
amenable to explicit computation after integration over the dual torus
parametrising the different holomorphic bundle structures within each
topological charge sector.  In particular, Jevicki's approach requires
computation of the residues of meromorphic functions instead of
functionals, and his programme can be carried out in the simplest
cases exhibiting the reappearance of instantons as resonances (poles)
in the one-loop effective action.  Remarkably, in spite of the absence
of regular unit charge instantons on the torus \cite{Braam:1989qk},
the contribution of this sector is nonzero in the resonance approach:
this we interpret as the effect of rough configurations near the
forbidden regular instanton.

At finite volume there are two regimes: one dominated by instantons
for low mass theories, $m^2 V \ll N$, and another regime where they
are are strongly suppressed, $m^2 V \gg N$. The second regime is the
relevant one for \cpn{} theories in the thermodynamic limit, but the
other regime is relevant for effects where the finite volume or space
topology play a leading r\^ole, like in the appearance of edge states
in the quantum Hall effect.

The structure of the article is as follows.  In section
\ref{sec:traditional}, the traditional large $N$ picture of \cpn{}
models is reviewed.  The $\theta$ dependence of \cpn{} models
formulated on the sphere is considered in section \ref{sec:sphere}.
The corresponding analysis for the case of the torus is performed in
section \ref{sec:torus}.  The consequences of this analysis are
discussed in section \ref{sec:discussion}.

\section{\label{sec:traditional}%
  The traditional picture of the $1/N$ expansion}

The traditional large $N$ picture of \cpn{} models was developed in
\cite{D'Adda:1978uc} and \cite{Witten:1979bc}. We shall now give a
brief account of it before the analysis in finite volume.

The large $N$ method is based in a saddle point approximation of the
partition function, defined on the infinite 2$d$ Euclidean plane,
after integration of the fundamental $\Psi$, $\Psi^\dagger$ fields
(taking values in $\mathbb{C}^N$, i.e., in representatives of
projective classes in \cpn).  We introduce the dummy $\mathrm{U} (1)$
gauge field
\begin{equation}\label{analisisn:dummyuone}
  A_\nu
=
 - \,
  \frac{ i }{ 2 }
  \left(
    \Psi^\dagger \partial_\mu \Psi
   -
    ( \partial_\mu \Psi )^\dagger \Psi
  \right) ,
\end{equation}
and a scalar field $\alpha (x)$ imposing the constraint $\Psi^\dagger
\Psi = 1$ at each point as a Lagrange multiplier.  Starting from the
full partition function at $\theta = 0$,
\begin{align}\label{analisisn:funciondeparticion}
\nonumber
  Z
&=
  \int
      \mathcal{D} \Psi \,
      \mathcal{D} \Psi^\dagger \,
      \mathcal{D} A_\mu \;
  \delta \!
        \left[
              \Psi^\dagger \Psi - 1
        \right]
  \exp \left\{
             - \,
             \frac{ N }{ 2 g_0^2 }
             \int_{\mathbb{R}^2} \mathrm{d}^2 x \,
             \lvert \mathrm{D}_\mu \Psi \rvert^2
       \right\}
\\
\nonumber
&=
  \int
      \mathcal{D} \Psi \,
      \mathcal{D} \Psi^\dagger \,
      \mathcal{D} A_\mu \,
      \mathcal{D} \alpha
\\
&
\qquad
 \times
  \exp \left\{
             - \,
             \frac{ N }{ 2 g_0^2 }
             \int_{\mathbb{R}^2} \mathrm{d}^2 x \,
             \lvert \mathrm{D}_\mu \Psi \rvert^2
            - \,
             \frac{ N }{ 2 g_0^2 }
             \int_{\mathbb{R}^2} \mathrm{d}^2 x \,
             \alpha (x)
             \left(
                   \Psi^\dagger \Psi - 1
             \right)
       \right\} ,
\end{align}
we perform the Gaussian integration over $\Psi$, $\Psi^\dagger$ to
obtain
\begin{equation}\label{analisisn:funciondeparticionaalfa}
  Z
=
  \int 
      \mathcal{D} A_\mu \,
      \mathcal{D} \alpha \,
  \mathrm{e}^{ - N S_\mathrm{eff} [ A_\mu, \, \alpha ] } ,
\end{equation}
where the effective action is
\begin{equation}\label{analisisn:accionefectivaaalfa}
  S_\mathrm{eff} [ A_\mu, \, \alpha ]
=
  \mathrm{Tr} \, \ln \left(
                - \mathrm{D}_\mu^2 - \alpha (x)
          \right)
 - \,
  \frac{ 1 }{ 2 g_0^2 }
  \int_{\mathbb{R}^2} \mathrm{d}^2 x \,
  \alpha (x) .
\end{equation}

The saddle point equations
\begin{align}
\label{analisisn:ecuacionpuntosillaalfa}
  \frac{
        \delta S_\mathrm{eff}
      }{
        \delta \alpha (x)
       }
&=
  \frac{
        1
      }{
        - \mathrm{D}_\mu^2 + \alpha
       }
  ( x, \, x )
 - \,
  \frac{ 1 }{ 2 g_0^2 }
=
  0 ,
\\
\label{analisisn:ecuacionpuntosillaamu}
  \frac{
        \delta S_\mathrm{eff}
      }{
        \delta A_\mu (x)
       }
&=
  2 i \,
  \frac{
        \mathrm{D}_\mu
      }{
        - \mathrm{D}_\mu^2 + \alpha
       }
  ( x, \, x )
=
  0 ,
\end{align}
can be solved within a renormalisation scheme to yield a saddle
configuration
\begin{equation}\label{extraextra}
  A_\mu
=
  0 ,
\quad
  \alpha
=
  m_{\mathrm{T}}^2
\equiv
  \mu^2
  \exp \left\{
             - \,
             \frac{ 2 \pi }{ g_R^2 ( \mu )}
       \right\} ,
\end{equation}
where $\mu$ is a mass scale and $g_R$ the corresponding renormalised
coupling.

Perturbation theory around the saddle configuration reveals a
dynamical system of an $N$-plet of charged scalars $\Psi$, with a
short range interaction due to the field $\alpha$, and electromagnetic
interaction due to the field $A_\mu$.  The latter develops an
effective kinetic term and couples to the scalars with effective
electric charge $e_{\mathrm{eff}} = \sqrt{ 12 \pi m_{\mathrm{T}}^2 \,
  / \, N}$.  Thus there is a confining Coulomb interaction between
scalars, and the spectrum at $\theta = 0$ consists of $(\Psi
\Psi^\dagger)$ bound states, with mass gap $2 m_{\mathrm{T}}$, living
in the adjoint representation of $\mathrm{SU} (N)$.

For $\theta \neq 0$, the topological term in the action is
\begin{equation}\label{analisisn:topologicalterm}
  - i \theta Q
 =
  - i
   \frac{ \theta }{ 2 \pi }
   \int \mathrm{d}^2 x \,
   F_{01} ,
\end{equation}
$Q$ being the magnetic flux associated with the $\mathrm{U} (1)$ field
$A_\mu$ and its field strength $F_{\mu\nu}$, or equivalently, the
topological charge of the field $\Psi$, which is an integer for smooth
finite-action configurations.  This term plays the r\^ole of an
external electric field in electrodynamics. Therefore, in this picture
of the \cpn{} models, its contribution to the vacuum energy density is
\begin{equation}\label{analisisn:densenergiacampoelectricofondo}
  \mathcal{E}_0 ( \theta )
=
  \frac{1}{2} \,
  e_{\mathrm{eff}}^2 \,
  \left( \frac{ \theta }{ 2 \pi } \right)^2
=
  \frac{ 3 }{ 2 \pi } \, \frac{ m_{\mathrm{T}}^2 \theta^2 }{ N } ,
\end{equation}
yielding a topological susceptibility
\begin{equation}\label{analisisn:suscepttopologica}
  \chi_t
=
  \left(
    \frac{
      \mathrm{d}^2 \mathcal{E}_0 ( \theta )
    }{
      \mathrm{d} \theta^2
    }
  \right)_{ \theta = 0 }
=
  \frac{ 3 m_{\mathrm{T}}^2 }{ \pi N } .
\end{equation}
This quadratic dependence is perturbative, i.e., it can be seen in
terms of Feynman diagrams.  Instanton effects, nonperturbative in
nature, were argued in \cite{Witten:1979bc} to be exponentially
suppressed in the $1/N$ expansion, and therefore irrelevant for the
physics of the \cpn{} models. However, we have seen that the level
crossing and first order phase transition at $\theta = \pi$ implied by
(\ref{analisisn:densenergiacampoelectricofondo}) and the requirement
of $2 \pi$-periodicity in $\theta$ are in contradiction with the nodal
arguments of \cite{Asorey:1998cz}.  We will next go over to a compact
space with the purpose of showing that this incompatibility stems from
the infinite volume starting point of the traditional $1/N$ analysis.

\section{\label{sec:sphere}%
  $1/N$ expansion on $\mathrm{S}^2$}

Clarifying the interplay of $N$ and the volume, and the effects of
taking these to infinity in different orders, requires the \cpn{}
models to be first formulated in a compact (Euclidean) space.

Schwab \cite{Schwab1, Schwab2} and M\"unster \cite{Munster:1982wg,
  Munster:1983sd} studied the $1/N$ expansion of \cpn{} models on
$\mathrm{S}^2$, and observed that the $k = 1$ contribution to the
partition function is dominated, for large $N$, by a saddle point
given by a rotationally invariant instanton (in the sense that global
$\mathrm{U} (N)$ transformations can be compensated by $\mathrm{O}
(3)$ rotations in Euclidean space).  The saddle point equations
\begin{align}\label{n_esfera:ecuacionespuntosilla}
\nonumber
  \frac{
        1
      }{
        - \mathrm{D}^\mu \mathrm{D}_\mu + \alpha
       }
  ( x, \, x )
&=
  \frac{ 1 }{ 2 g_0^2 } ,
\\
  \frac{
        \mathrm{D}_\nu
      }{
        - \mathrm{D}^\mu \mathrm{D}_\mu + \alpha
       }
  ( x, \, x )
&=
  0 ,
\end{align}
admit for $N > \lvert k \rvert$, in a uniform topological charge
density background, solutions with constant $\alpha ( x )$.  Indeed,
the second equation holds due to parity. The first equation states
rotation invariance of the propagator $G (x, \, x)$ of a particle with
mass $\sqrt{\alpha}$.  It is easy to show that $G (x, \, y)$ depends
only on the geodesic distance between $x$ and $y$, and therefore the
first equation has solutions with constant $\alpha$.

\subsection{\label{subsec:sphere:effaction}%
  The effective action on the sphere}

Thus, we begin \cite{Aguado:phdthesis, aaga} with a spherical
spacetime of radius $R$ and volume $V = 4 \pi R^2$, and the action of
the \cpn{} model on a background of topological charge $k$,
\begin{equation}\label{n_esfera:accionsectork}
  S_k
=
  -\,
  \frac{ N }{ 2 g_0^2 }
  \int \Psi^\dagger \Delta_k \Psi
 +
  \frac{ N }{ 2 g_0^2 }
  \int m^2
  \left( \Psi^\dagger \Psi - 1 \right) ,
\end{equation}
where integration implies the measure $\mathrm{d}^2 x \, \sqrt{g}$,
and $\Delta_k$ is the covariant Laplacian in the background chosen.
The magnetic flux for the composite $\mathrm{U} (1)$ field is
quantised,
\begin{equation}\label{n_esfera:condiciondedirac}
  \Phi_B
=
  4 \pi R^2 B
=
  2 \pi k ,
\quad
  k \in \mathbb{Z} .
\end{equation}
We rewrite the constant saddle point value of the $\alpha$ field as
$m^2$, variable still to be integrated upon.

Integrating out $\Psi$, $\Psi^\dagger$ yields the functional
determinant of the operator $- \Delta_k + m^2$, which is computed in
the $\zeta$ function renormalisation scheme at energy scale $\mu$.
Discarding unessential factors,
\begin{equation}\label{n_esfera:accionefectivacondeterminante}
  Z_k
=
  \int \mathrm{d} m^2 \,
  \mathrm{e}^{ - N S_k^{\rm eff} } ,
\end{equation}
with effective action
\begin{align}\label{n_esfera:accionefectivalndet}
\nonumber
  S_k^{\rm eff}
&=
  \ln \det
        \left(
              \frac{ -1 }{ \mu^2 }
              \Delta_k
              +
              \frac{ m^2 }{ \mu^2 }
        \right)
 - \,
  \frac{ 4 \pi R^2 }{ 2 g_0^2 } m^2
\\
&\equiv
  \ln \det \mathcal{A}
  -
  \frac{ 4 \pi R^2 }{ 2 g_0^2 } m^2 .
\end{align}
The eigenvalues of $\mathcal{A}$ are
\begin{align}\label{n_esfera:autovaloresdea}
\nonumber
  \lambda_n
&=
  \frac{ 1 }{ \mu^2 R^2 }
  \left[
        \left( n + \frac{\lvert k \rvert}{2} + 1 \right)
        \left( n + \frac{\lvert k \rvert}{2} \right)
        - \frac{ k^2 }{ 4 }
        + m^2 R^2
    \right]
\\
&\equiv
  \frac{ {\tilde \lambda}_n }{ \mu^2 R^2 } ,
\qquad
  n = 0, \, 1, \, 2, \, \ldots ,
\end{align}
with degeneracy $d_n = 2 n + \lvert k \rvert + 1$.

We use the $\zeta$-function definition of the determinant, equivalent
to a renormalisation at scale $\mu$:
\begin{align}\label{n_esfera:prescripcionzeta}
\nonumber
  \ln \det{}_\zeta \mathcal{A}
&=
  \sum_{n=0}^\infty
  d_n \ln \lambda_n
=
  \sum_{n=0}^\infty
  d_n \ln {\tilde\lambda}_n  
 -
  \left( \sum_{n=0}^\infty d_n \right)
  \ln ( \mu^2 R^2 )
\\
&\rightarrow
 - \,
  \zeta'_{\tilde{\mathcal{A}}} (0)
 - \,
  \zeta_{\tilde{\mathcal{A}}} (0)
  \ln ( \mu^2 R^2 ) ,
\end{align}
$\zeta_{\tilde{\mathcal{A}}}$ being the $\zeta$ function associated
with the operator $\tilde{\mathcal{A}} \equiv \mu^2 R^2 \mathcal{A}$,
i.e. the analytic continuation of
\begin{equation}\label{n_esfera:deffuncionzeta}
  \zeta_{\tilde{\mathcal{A}}} (s)
\equiv
  \sum_{n=0}^\infty
  \frac{ d_n }{ {\tilde\lambda}_n^s } ,
\qquad
  ( {\rm Re}\,s > 1 )
\end{equation}
for all complex $s \neq 1$.

{}From the small $s$ expansion of (\ref{n_esfera:deffuncionzeta}), we
obtain the effective action for topological sector $k$:
\begin{align}\label{n_esfera:accionrenormalizada}
\nonumber
  S^{\mathrm{eff}}_k
=&
  - \zeta'_{\tilde{\mathcal{A}}} (0)
  - \zeta_{\tilde{\mathcal{A}}} (0)
    \ln ( \mu^2 R^2 )
  - \frac{4 \pi R^2}{2 g_R^2} m^2
\\
\nonumber
=& \;
  2 \left(
          \frac{k^2}{4} + \frac{1}{4} - m^2 R^2
    \right)
  + \left( m^2 R^2 - \frac{1}{3} \right)
    \ln ( \mu^2 R^2 )
  - \,
    \frac{4 \pi R^2}{2 g_R^2} m^2
\\
\nonumber
&
 + 2 \sqrt{ \frac{k^2}{4} + \frac{1}{4} - m^2 R^2 } \,
  \ln \frac{
            \Gamma
              \left(
                \frac{\lvert k \rvert+1}{2}
                + \sqrt{ \frac{k^2}{4} + \frac{1}{4} - m^2 R^2 }
              \right)
          }{
            \Gamma
              \left(
                \frac{\lvert k \rvert+1}{2}
                - \sqrt{ \frac{k^2}{4} + \frac{1}{4} - m^2 R^2 }
              \right)
           }
\\
\nonumber
&
 -
  2 \zeta^\prime_H\!\!
       \left(
          - 1 ;
          \frac{\lvert k \rvert+1}{2}
          + \sqrt{ \frac{k^2}{4} + \frac{1}{4} - m^2 R^2 }
        \right)
\\
&
 -
  2 \zeta^\prime_H\!\!
       \left(
          - 1 ;
          \frac{\lvert k \rvert+1}{2}
          - \sqrt{ \frac{k^2}{4} + \frac{1}{4} - m^2 R^2 }
        \right) .
\end{align}
Here we have used the Hurwitz zeta function $\zeta_H ( s; v )$,
defined by analytical continuation to all $s \neq 1$ of
\begin{equation}\label{n_esfera:hurwitzzeta}
  \zeta_H ( s; v )
=
  \sum_{ n = 0 }^\infty ( n + v )^{-s}
=
  \frac{ 1 }{ \Gamma (s) }
  \int_0^\infty \mathrm{d} t \,
  \frac{
    t^{ s - 1 } \mathrm{e}^{- v t}
  }{
    1 - \mathrm{e}^{-t}
  } ,
\quad
  \mathrm{Re} \, s > 1,
\end{equation}
and its derivative $\zeta^\prime_H ( s; v )$ with respect to $s$.
Function (\ref{n_esfera:accionrenormalizada}) is defined for all
complex values of $m^2 R^2$, bar isolated singularities.

\subsection{\label{subsec:sphere:compeffaction}%
  Zeros and saddle points of the effective action}

In order to compute $Z_k$, integration over $m^2$ is still to be
performed, through imaginary values in order to ensure convergence.

\begin{figure}
  \centering
  \psfrag{mr}{$m^2 R^2$}
  \psfrag{seff}{$S_k^{\mathrm{eff}}$}
  \includegraphics{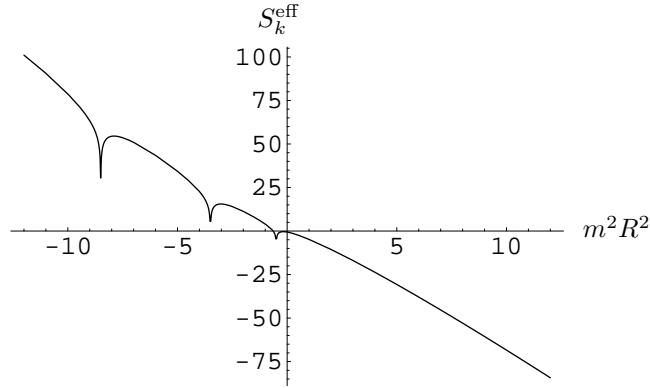}
  \caption{Effective action in the $k = 1$ sector on the sphere,
    for real $m^2 R^2$.}
  \label{n_esfera:fig:sefesfera}
\end{figure}
Let us study the behaviour of the effective action for \textsl{real}
$m^2 R^2$ (see figure \ref{n_esfera:fig:sefesfera}).  To begin with,
the integrand of $Z_k$ has $N ( 2n + \lvert k \rvert + 1 )$-fold poles
at
\begin{equation}\label{n_esfera:lospolosdondedondeestan}
  m^2 R^2
=
  p_n
=
 -
  \left( n + \frac{\lvert k \rvert}{2} + 1 \right)
  \left( n + \frac{\lvert k \rvert}{2} \right)
 +
  \frac{ k^2 }{ 4 } ,
\quad
  n = 0, \, 1, \, 2, \, \ldots,
\end{equation}
reproducing the eigenvalues and degeneracies of $- \Delta_k + m^2$. In
this sense, in the way predicted by Jevicki, the partition function
can be computed by deforming the integration curve so as to surround
the poles, and summing the residues at each of them.  However, the
problem of computing and summing the residues is too difficult to be
tackled analytically, although the previous formulae can be used in a
numerical approach (more progress can be made analytically in the case
of the torus, as will be seen in next section).

Alternatively, we can use the saddle point method. The zeros of the
derivative,
\begin{align}\label{n_esfera:derivadasefectiva}
\nonumber
&
  \frac{
        \mathrm{d} S^{\mathrm{eff}}_k
      }{
        \mathrm{d} (m^2 R^2)
       }
=
  \ln ( \mu^2 R^2 )
 - \,
  \frac{ 4 \pi }{ 2 g_R^2 }
\\
&
\quad
 - \,
  \psi \left(
             \frac{\lvert k \rvert+1}{2} 
             + \sqrt{ \frac{k^2+1}{4} - m^2 R^2 }
       \right)
 - \,
  \psi \left(
             \frac{\lvert k \rvert+1}{2} 
             - \sqrt{ \frac{k^2+1}{4} - m^2 R^2 }
       \right) ,
\end{align}
alternate with the poles in the real $m^2 R^2$ axis, as seen in figure
\ref{n_esfera:fig:sefesfera} (here $\psi (z) = \Gamma' (z) / \Gamma
(z)$ is the digamma function).  There is a unique saddle point $s_0$
to the right of the first pole $p_0 = - \, \frac{\lvert k \rvert}{2}$,
which we assume to be dominant.

The partition function of sector $k$ in the saddle point approximation
is
\begin{equation}\label{n_esfera:aproxpuntosillascero}
  Z_k^{(s_0)}
=
  \frac{ 1 }{ R^2 } \,
  \mathrm{e}^{ - N S_k^{\mathrm{eff}} (s_0) } \,
  \sqrt{
    \frac{
      2 \pi
    }{
      N \lvert S_k^{\mathrm{eff}}{}''(s_0) \rvert
    }
  } ,
\end{equation}
up to quadratic order. Explicit results can be obtained for large $m^2
R^2$, in which region the effective action can be expanded as
\begin{align}\label{n_esfera:sefectivagranmr}
\nonumber
  S^{\mathrm{eff}}_k
&=
 - \left( m^2 R^2 - \frac{1}{3} \right)
  \ln \frac{ m^2 }{ \mu^2 }
 - \frac{ 4 \pi R^2 }{ 2 g_R^2 } m^2
 + m^2 R^2
\\
&
\qquad
 + \left( \frac{k^2}{24} - \, \frac{1}{15} \right)
  \frac{ 1 }{ m^2 R^2 } 
 + \left( \frac{k^2}{40} - \, \frac{4}{315} \right)
  \frac{ 1 }{ m^4 R^4 } 
 + {\cal O} \left( m^{-6} R^{-6} \right) ,
\end{align}
and the saddle point is found to be
\begin{align}\label{n_esfera:ecuaciondegapcorregida}
\nonumber
  m_k^2 R^2
&=
  m_{\mathrm{T}}^2 R^2
 + \frac{1}{3}
 - \left( \frac{k^2}{24} - \, \frac{1}{90} \right)
  \frac{ 1 }{ m_{\mathrm{T}}^2 R^2 }
\\
&
\qquad
 - \left( \frac{k^2}{45} - \, \frac{16}{2835} \right)
  \frac{ 1 }{ m_{\mathrm{T}}^4 R^4 }
 + {\cal O} ( m^{-6} R^{-6} ) ,
\end{align}
where $m_{\mathrm{T}}^2 = \mu^2 \exp \left\{ - \, 2 \pi / g_R^2
\right\}$ is the infinite volume saddle point in (\ref{extraextra}).

The total partition function after summing all topological sectors is
\begin{align}\label{n_esfera:zetatotalsaddle}
\nonumber
&
  Z^{(s_0)} (\theta)
=
  \sum_{ k \in \mathbb{Z} }
  Z_k^{(s_0)} \,
  \mathrm{e}^{- i k \theta}
\\
\nonumber
&=
  \sqrt{ \frac{2 \pi}{N} } \,
  \frac{ m_{\mathrm{T}} }{ R } \,
  \exp \left\{
         \frac{2 \pi N}{3 g_R^2} 
         - N m_{\mathrm{T}}^2 R^2
         + \frac{ N }{ 90 m_{\mathrm{T}}^2 R^2 }
       \right\}
\\
\nonumber
&\qquad
 \times
  \sum_{ k \in \mathbb{Z} }
  \exp \left\{
        - \, \frac{ N k^2 }{ 24 m_{\mathrm{T}}^2 R^2 }
        - i k \theta
       \right\} \,
  \left( 1 +
        \mathcal{O} \left( m_{\mathrm{T}}^{-4} R^{-4} \right)
  \right)
\\
\nonumber
&\equiv
  \sqrt{ \frac{2 \pi}{N} } \,
  \frac{ m_{\mathrm{T}} }{ R } \,
  \exp \left\{
         \frac{2 \pi N}{3 g_R^2}
         - N m_{\mathrm{T}}^2 R^2
         + \frac{ N }{ 90 m_{\mathrm{T}}^2 R^2 }
       \right\}
\\
&\qquad
 \times
  \vartheta_3
    \left(
      \frac{ \theta }{ 2 \pi }
    \right\vert \left.
      \frac{ i N }{ 24 \pi m_{\mathrm{T}}^2 R^2 }
    \right)
  \left( 1 +
        \mathcal{O} \left( m_{\mathrm{T}}^{-4} R^{-4} \right)
  \right) ,
\end{align}
where the last equation uses Jacobi's $\vartheta_3$ function:
\begin{equation}\label{n_toro:varthetatres}
  \vartheta_3 ( z \vert \tau )
=
  \sum_{ n \in \mathbb{Z} }
  \mathrm{e}^{ i \pi \tau n^2 } 
  \mathrm{e}^{ i 2 \pi n z } .
\end{equation}

Two asymptotic regimes for (\ref{n_esfera:zetatotalsaddle}) can be
analysed. For $N \gg m_{\mathrm{T}}^2 R^2$, the sum therein can be
truncated, keeping just the $k = -1, \, 0, \, 1$ sectors.  Then the
vacuum energy density has a typical dilute instanton gas $\theta$
dependence,
\begin{align}\label{n_esfera:densevaciongrande}
\nonumber
  \mathcal{E}_0 (\theta) - \mathcal{E}_0 (0)
&=
 - \,\frac{ 1 }{ 4 \pi R^2 } \,
  \ln \frac{ Z^{(s_0)} (\theta) }{ Z^{(s_0)} (0) }
\\
&\approx
  \frac{ 1 }{ 4 \pi R^2 } \,
  \exp \left\{ - \, \frac{N}{24 m_{\mathrm{T}}^2 R^2} \right\} \,
  ( 1 - \cos \theta ) .
\end{align}

But if $m_{\mathrm{T}}^2 R^2 \gg N$, using the Poisson resummation
formula for the $\theta$ function in (\ref{n_esfera:zetatotalsaddle})
and keeping the dominant term in the dual sum, we have
\begin{equation}\label{n_esfera:zetatotalpoissonaprox}
  Z^{(s_0)} ( \theta )
\approx
  \frac{ 4 \sqrt{3} \, \pi m_{\mathrm{T}}^2 }{ N } \,
  \exp \left\{
         \frac{2 \pi N}{3 g_R^2} \,
         - N m_{\mathrm{T}}^2 R^2
         + \frac{ N }{ 90 m_{\mathrm{T}}^2 R^2 }
         - \, \frac{ 6 m_{\mathrm{T}}^2 R^2 }{ N } \,
         {\tilde{\theta}}^2
       \right\} ,
\end{equation}
$\tilde{\theta}$ being the angle in $( - \pi, \, + \pi ]$ differing
  from $\theta$ by an integer.  The corresponding vacuum energy
  density coincides with the traditional large $N$ prediction:
\begin{equation}\label{n_esfera:densenergiavaciotruncagranv}
  \mathcal{E}_0^{(s_0)} ( \theta )
 -
  \mathcal{E}_0^{(s_0)} ( 0 )
\approx
  \frac{ 3 m_{\mathrm{T}}^2 }{ 2 \pi N } \, {\tilde{\theta}}^2 ,
\end{equation}
which is periodic in $\theta$ and undergoes first order phase
transitions with level crossing at $\theta = (2 \ell + 1) \pi$, $\ell
\in \mathbb{Z}$.

Before commenting on these two different limiting procedures, let us
perform the same analysis on the torus \cite{Aguado:phdthesis, aaga}.

\section{\label{sec:torus}%
  $1/N$ expansion on $\mathrm{T}^2$}

We consider a toric spacetime of linear size $L$ and spacetime volume
$V = L^2$.  Functional integration over the fields of the \cpn{} model
on the torus involves an additional variable, the complex coordinate
$u \in {\hat{\mathrm{T}}}^2$ in the dual torus parametrising the
different holomorphic bundle structures associated with the complex
line bundle $E_k ( \mathrm{T}^2, \, \mathbb{C} )$
\cite{Asorey:1998wu}.

In this case, the effective action $S_u^{\mathrm{eff}} [A_\mu, \,
  \alpha]$ resulting from integration of the $\Psi$, $\Psi^\dagger$
fields does not have saddle points.  Specifically, the saddle point
equations
\begin{align}\label{n_toro:puntosillatoro}
\nonumber
  \frac{
        1
      }{
        - \mathrm{D}_\mu^2 + \alpha
       }
  ( x, \, x )
&=
  \frac{ 1 }{ 2 g_0^2 } ,
\\
  \frac{
        \mathrm{D}_\mu
      }{
        - \mathrm{D}_\mu^2 + \alpha
       }
  ( x, \, x )
&=
  0 ,
\end{align}
do not have solutions with constant $\alpha$ and topological charge
density.

\subsection{\label{subsec:torus:effaction}%
  Quantum saddle points and the effective action on $\mathrm{T}^2$}

The arguments used for the sphere can, nevertheless, be adapted to the
torus, generalising the saddle point method.  By integrating over $u$
(i.e., averaging over $u$), the irregularities of the saddle point
configurations are swept off \cite{Aguado:phdthesis, aaga}.  Upon
integration over $u$, a reduced effective action $S_{ \mathrm{red} }$
obtains,
\begin{equation}\label{n_toro:accionefectivaintegrandou}
  \exp \left\{ - S_{\mathrm{red}} [ A_\mu, \, \alpha, \, N ] \right\}
=
  \int_{ {\hat{\mathrm{T}}}^2 } \mathrm{d}^2 u \,
  \mathrm{e}^{ - N S_u^{\mathrm{eff}} [ A_\mu, \, \alpha ] } ,
\end{equation}
which can be argued to be dominated by constant topological density in
the large $N$ limit. The generalised saddle point equations
\begin{align}\label{n_toro:generalizarpuntosilla}
\nonumber
  \frac{ \delta }{ \delta A_\mu } \,
  \frac{ \partial }{ \partial N } \,
  S_{\mathrm{red}} [ A_\mu, \, \alpha, \, N ]
&=
  0 ,
\\
  \frac{ \delta }{ \delta \alpha } \,
  \frac{ \partial }{ \partial N } \,
  S_{\mathrm{red}} [ A_\mu, \, \alpha, \, N ]
&=
  0 ,
\end{align}
hold in this case because they remain finite as $N \rightarrow
\infty$. Their solutions we call \textsl{quantum saddle points}.

To compute the effective action in the sectors of nonzero topological
charge, the $\zeta$ function method is used, renormalising at energy
scale $\mu$:
\begin{align}\label{n_toro:accionefectivalndet}
\nonumber
  S_k^{\rm eff}
&=
  \ln \det
        \left(
              \frac{ -1 }{ \mu^2 }
              \Delta_k
              +
              \frac{ m^2 }{ \mu^2 }
        \right)
 - \,
  \frac{ m^2 L^2 }{ 2 g_0^2 }
\\
&\equiv
 - \zeta'_{\mathcal{B}} (0)
 - \zeta_{\mathcal{B}} (0)
  \ln \frac{ \mu^2 L^2 }{ 4 \pi \lvert k \rvert }
 - \,
  \frac{ m^2 L^2 }{ 2 g_0^2 } .
\end{align}
The spectrum of the Laplacian on the torus, in a background with
uniformly distributed topological charge $2 \pi k \neq 0$, is
independent of the holonomies $u$ and consists of Landau levels $- 2
\omega ( n + 1/2 )$, $n = 0, \, 1, \, \ldots$, with $\omega = \lvert B
\rvert = \frac{2 \pi \lvert k \rvert}{L^2}$, where $L$ is the linear
size of the torus.  These levels have $\lvert k \rvert$-fold
degeneracy \cite{Aguado:2000aw}. The zeta function for operator
$\mathcal{B} = \big( - \, L^2 \Delta_k + m^2 L^2 \big) / ( 4 \pi
\lvert k \rvert )$ is
\begin{align}\label{n_toro:funcionzetadebe}
\nonumber
  \zeta_{\mathcal{B}} (s)
&=
  \sum_{n=0}^\infty \,
  \lvert k \rvert \,
  \left( 
    n + \frac{m^2 L^2}{4 \pi \lvert k \rvert} + \frac{1}{2} 
  \right)^{-s}
=
  \lvert k \rvert \,
  \zeta_H \left(
            s ; \,
            \frac{m^2 L^2}{4 \pi \lvert k \rvert}
            + \frac{1}{2}
          \right)
\\
&=
 - \lvert k \rvert \, \frac{m^2 L^2}{4 \pi \lvert k \rvert}
 + s \, \lvert k \rvert \,
  \ln \left\{
        \frac{ 1 }{ \sqrt{ 2 \pi } } \,
        \Gamma \left(
                 \frac{m^2 L^2}{4 \pi \lvert k \rvert}
                 + \frac{1}{2}
               \right)
      \right\}
 + \mathcal{O} ( s^2 ) ,
\end{align}
yielding the effective action (see \cite{Blau:1991iz})
\begin{align}\label{n_toro:accionefectiva}
\nonumber
  S_k^{\mathrm{eff}}
&=
 - \,
  \frac{ m^2 L^2 }{ 4 \pi } \,
  \left\{
    \frac{ 2 \pi }{ g_R^2 }
   +
    \ln \frac{ 4 \pi \lvert k \rvert }{ \mu^2 L^2 }
  \right\}
 - \,
  \lvert k \rvert \,
  \ln \left\{
        \frac{ 1 }{ \sqrt{ 2 \pi } } \,
        \Gamma \left(
                 \frac{m^2 L^2}{4 \pi \lvert k \rvert}
                 + \frac{1}{2}
               \right)
      \right\}
\\
&=
 - \,
  \frac{ m^2 L^2 }{ 4 \pi } \,
    \ln \frac{ 4 \pi \lvert k \rvert }{ m_{\mathrm{T}}^2 L^2 } \,
 - \,
  \lvert k \rvert \,
  \ln \left\{
        \frac{ 1 }{ \sqrt{ 2 \pi } } \,
        \Gamma \left(
                 \frac{m^2 L^2}{4 \pi \lvert k \rvert}
                 + \frac{1}{2}
               \right)
      \right\}
\end{align}
where $m^2$ is the (constant) saddle point value of the $\alpha$
field, and $g_R = g_R ( \mu )$ is the renormalised coupling at scale
$\mu$.  In the last equation, $m_{\mathrm{T}}^2 = \mu^2 \exp \big\{ -
\, 2\pi / g_R^2(\mu) \big\}$ stands for the large $N$ dynamically
generated mass at infinite volume.

Expression (\ref{n_toro:accionefectiva}) can be checked to coincide
with the dominant term in a large volume, constant $B$ expansion of
the corresponding effective action
(\ref{n_esfera:accionrenormalizada}) for the sphere:
\begin{equation}\label{n_toro:accionktorolimiteesfera}
  S_{\mathrm{ef},k}^{\mathrm{S^2}}
\stackrel{{V \rightarrow \infty}}{\longrightarrow}
  S_{\mathrm{ef},k}^{\mathrm{T^2}}
 +
  \mathcal{O} ( V^0 )
\qquad
  ( B = \mathrm{const} ) .
\end{equation}

The contribution of topological sector $k$ to the partition function
on the torus now depends only on $m^2$, all other fields having been
integrated out:
\begin{equation}\label{n_toro:funciondeparticionconm}
  Z_k
=
  \int \mathrm{d} m^2 \,
  \mathrm{e}^{ - N S_k^{\mathrm{ef}} }.
\end{equation}
As in the previous sections, the integration is performed through
imaginary values of $m^2$ to guarantee convergence.

In order to make the functional dependences in some the following
expressions clear, it is useful to define dimensionless variables
\begin{equation}\label{n_toro:masasreescaladas}
  y
=
  \frac{ m^2 L^2 }{ 4 \pi \lvert k \rvert } ,
\quad
  y_0
=
  \frac{ m_{\mathrm{T}}^2 L^2 }{ 4 \pi \lvert k \rvert } .
\end{equation}

\begin{figure}
  \centering
  \psfrag{y}{$y$}
  \psfrag{seff}{$\tilde{S} (y)$}
  \includegraphics{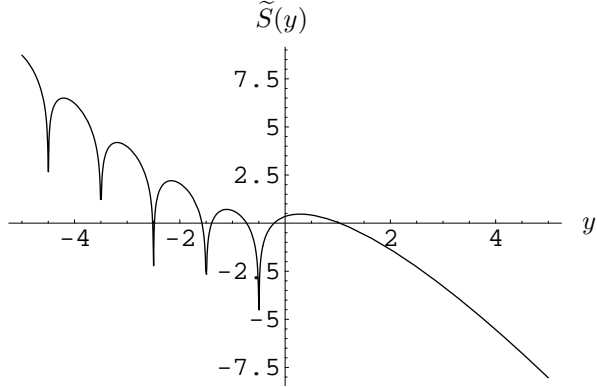}
  \caption{Effective action $\tilde{S} (y)$ for the torus, with $y \in
    \mathbb{R}$.}
  \label{n_toro:fig:sdeytoro}
\end{figure}

Then the $k$-sector partition function is
\begin{equation}\label{n_toro:funciondeparticioncony}
  Z_k
=
  \frac{ 4 \pi \lvert k \rvert }{ L^2 }
  \int \mathrm{d} y \,
  \left(
        \frac{
              \Gamma \left( y + \frac{1}{2} \right)
            }{
              \sqrt{ 2 \pi }
             }
  \right)^{ \!\! N \lvert k \rvert }
  \mathrm{e}^{ - N \lvert k \rvert y \ln y_0 }
=
  \frac{ 4 \pi \lvert k \rvert }{ L^2 }
  \int \mathrm{d} y \;
  \mathrm{e}^{ - N \lvert k \rvert \tilde{S} (y) } ,
\end{equation}
where the function
\begin{equation}\label{n_toro:exponentedefparticioncony}
  \tilde{S} (y)
=
  \frac{ S^{\textrm{eff}}_k }{ \lvert k \rvert }
=
  y \ln y_0
 -
  \ln \frac{
            \Gamma \left( y + 1/2 \right)
          }{
            \sqrt{ 2 \pi }
           }
\end{equation}
is defined for all complex values of $y$, except for a series of poles
of the integrand of $Z_k$, as can be seen in figure
\ref{n_toro:fig:sdeytoro}.

For small values of $y$, that is, when $\lvert k \rvert \ll m^2 L^2$,
the exponent simplifies:
\begin{equation}\label{n_toro:exponenteyengrany}
  \tilde{S} ( y )
=
  - y \ln \frac{ y }{ y_0 }
 +
  y
 +
  \frac{ 1 }{ 24 \, y }
 - \,
  \frac{ 7 }{ 2880 \, y^3}
 +
  \mathcal{O} \left( y^{-4} \right) ,
\end{equation}
meaning that the effective action has an expansion in powers of the
topological number $k$ where the first nontrivial term is quadratic:
\begin{equation}\label{n_toro:exponenteyengranyphysical}
  S^{\mathrm{eff}}_k
=
 - \,
  \frac{ m^2 L^2 }{ 4 \pi }
  \ln \frac{ m^2 }{ m_{\mathrm{T}}^2 }
 +
  \frac{ m^2 L^2 }{ 4 \pi }
 +
  \frac{ \pi k^2 }{ 6 \, m^2 L^2 }
 + \,
  \mathcal{O} \left( \frac{ k^4 }{ m^6 \, L^6 } \right) .
\end{equation}

In the opposite limit, when $\lvert k \rvert \gg m^2 L^2$, that is,
for large $y$,
\begin{equation}\label{n_toro:exponenteyenpequenyoy}
  \tilde{S} ( y )
=
  \frac{ \ln 2 }{ 2 }
 + 
  \big\{
        \ln y_0
       -
        \psi ( 1/2 )
  \big\} \,
  y
 - \,
  \frac{ \pi^2 }{ 4 } \,
  y^2
 - \,
  \frac{1}{6} \,
  \psi'' \! \left( 1/2 \right)
  y^3
 - \,
  \frac{ \pi^2 }{ 24 } \,
  y^4
 +
  \mathcal{O} ( y^5 ) .
\end{equation}
Written in terms of the effective action, we see that the leading term
is linear in the absolute value of the topological number:
\begin{equation}\label{n_toro:exponenteyenpequenyoyphysical}
  S^{\mathrm{eff}}_k
=
  \frac{ \ln 2 }{ 2 } \,
  \lvert k \rvert
 + 
  \frac{ m^2 L^2 }{ 4 \pi }
  \left\{
    \ln \frac{ m_{\mathrm{T}}^2 L^2 }{ 4 \pi \lvert k \rvert }
   -
    \psi ( 1/2 )
  \right\}
 +
  \mathcal{O}
  \left( \frac{ m^4 L^4 }{ \lvert k \rvert } \right) .
\end{equation}
Notice the change of asymptotic behaviour of the effective action
$S^{\mathrm{eff}}_k$ in the different topological sectors.  According
to equations (\ref{n_toro:exponenteyengranyphysical}) and
(\ref{n_toro:exponenteyenpequenyoyphysical}), for small values of the
topological charge, $\lvert k \rvert < m^2 V$, the effective action is
quadratic in $k$, whereas for large topological charges its leading
term is linear in $\lvert k \rvert$ \cite{aa2009}. This change of
asymptotic behaviour has important physical consequences.

\subsection{\label{subsec:torus:compeffaction}%
  Zeros and saddle points on the torus}

Extrema of $\tilde{S} ( y )$ allow us to perform a saddle point
approximation by deforming the integration contour so that it passes
through the dominant extremum.  The exponent in the $Z_k$ integral has
an overall $N \lvert k \rvert$ factor, therefore the saddle point
approximation is a large $N \lvert k \rvert$ expansion, and results in
terms of $y$ are general for all $k \neq 0$ sectors.

Poles of the integrand of $Z_k$ give us a chance of testing Jevicki's
proposal, since functional integration has been reduced to integration
along a path in the complex plane.  The integration contour must be
deformed so that it surrounds each pole.  In spite of the fact that
various fields have been integrated out in the effective action we are
working with, we shall see that instantons reappear in these poles.

\begin{figure}
  \centering
  \includegraphics{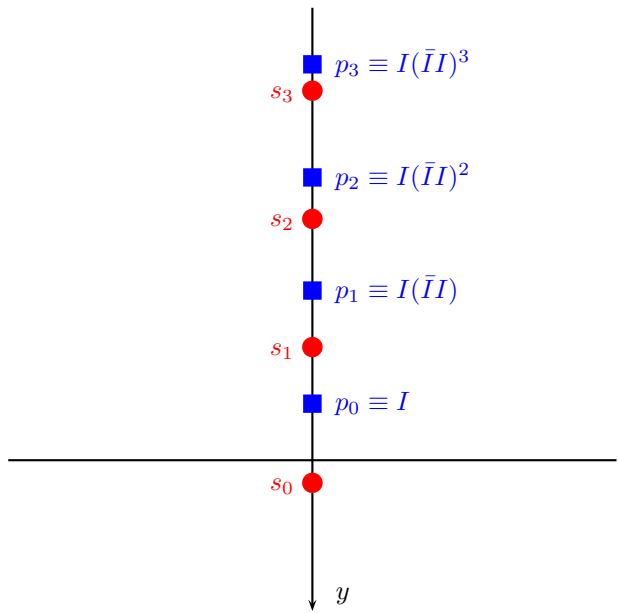}
  \caption{Structure of poles $p_n$ of the effective action (squares)
    and saddle points $s_n$ of the integrand of the partition function
    (circles) on the torus, in the complex plane of the variable $y =
    m_{\mathrm{T}}^2 L^2 / ( 4 \pi \lvert k \rvert )$.  Note that
    zeros and poles tend to coalesce as $n$ grows.}
  \label{n_toro:fig:cerosypolos}
\end{figure}

The structure of saddle points of $\tilde{S}_k$ and poles of the
integrand of $Z_k$ is represented in figure
\ref{n_toro:fig:cerosypolos}.
%

\subsubsection{\label{subsubsec:torus:poles}%
  Poles and Jevicki's approach}

Let us consider the poles.  There is an infinite series of $N \lvert k
\rvert$-fold poles at values $m_n$ of $m$ such that
\begin{equation}\label{n_toro:polos}
  y
=
  \frac{ m_n^2 L^2 }{ 4 \pi \lvert k \rvert }
\equiv
  p_n
=
  - \left( n + \frac{1}{2} \right),
\quad
  n = 0, \, 1, \, 2, \, \ldots .
\end{equation}
Their multiplicity is equal to the complex dimension of the moduli
space of charge $k$ instantons in the \cpn{} model (see
\cite{Aguado:2001xg}).

The partition function can be written as a sum over residues \`a la
Jevicki,
\begin{align}\label{n_toro:accionsumaderesiduos}
\nonumber
  Z_k
&=
  \frac{ 8 \pi^2 \lvert k \rvert }{ L^2 }  
  \sum_{ n = 0 }^\infty
  \Res { y \rightarrow p_n }
  \left(
        \frac{
              \Gamma \left( y + \frac{1}{2} \right)
            }{
              \sqrt{ 2 \pi }
             }
  \right)^{ \!\! N \lvert k \rvert }
  \mathrm{e}^{ - N \lvert k \rvert y \ln y_0 }
\\
&=
  \frac{ 8 \pi^2 \lvert k \rvert }{ L^2 } \,
  \left(
    \frac{ y_0 }{ 2 \pi }
  \right)^{ \frac{N \lvert k \rvert}{2} }
  \sum_{ n = 0 }^\infty
  \mathrm{e}^{ n N \lvert k \rvert \ln y_0 }
  \Res{ \varepsilon \rightarrow 0 }
  \left\{
         y_0^{ - \varepsilon } \,
         \Gamma ( - n + \varepsilon )
  \right\}^{ N \lvert k \rvert } .
\end{align}
There is no difficulty in computing and summing the residues for the
first cases, $N \lvert k \rvert = 1$ and $N \lvert k \rvert = 2$:
\begin{equation}\label{n_toro:sumaresiduosnkuno}
  Z_{ N \lvert k \rvert = 1 }
=
  \sqrt{2} \,
  \frac{ 2 \pi m_{\mathrm{T}} }{ L } \,
  \exp \left\{
             - \,
             \frac{ m_{\mathrm{T}}^2 L^2 }{ 4 \pi }
       \right\}
\end{equation}
and
\begin{equation}\label{n_toro:sumaresiduosnkdos}
  Z_{ N \lvert k \rvert = 2 }
=
  m_{\mathrm{T}}^2 \,
  \mathrm{K}_0
    \left(
      \frac{ 2 m_{\mathrm{T}}^2 L^2 }{ 4 \pi \lvert k \rvert }
    \right)
\end{equation}
(where $\mathrm{K}_0$ is a modified Bessel function), but the
partition function for higher values of $N \lvert k \rvert$ turns out
to be more difficult to compute.  From the expansion of
(\ref{n_toro:accionsumaderesiduos}), we can write it as
\begin{equation}\label{n_toro:sumandoespero}
  Z_k
=
  \frac{ 8 \pi^2 \lvert k \rvert }{ L^2 } \,
  \left(
        \frac{ y_0 }{ \sqrt{ 2 \pi } }
  \right)^{ \frac{N \lvert k \rvert}{2} }
  \sum_{ n = 0 }^\infty
  \left(
        \frac{ (-y_0)^n }{ n! }
  \right)^{ \! N \lvert k \rvert }
  T_{ N \lvert k \rvert - 1, \, n } (y_0) .
\end{equation}
The function $T_{ R, \, n } (y_0)$ is given by
\begin{equation}\label{n_toro:tresiduo}
  T_{ R, \, n } (y_0)
=
  \sum_{r,s,t=0}^\infty
  \delta_{r+s+t, \, R} \,
  a_r \, b_s \, c_{t,n},
\end{equation}
with coefficients defined by the expansions
\begin{align}\label{n_toro:abcdef}
\nonumber
  y_0^{-\varepsilon}
&=
  \sum_{r=0}^\infty a_r \varepsilon^r ,
\\
\nonumber
  \frac{ \pi \varepsilon }{ \sin \pi \varepsilon }
&\sim
  \sum_{s=0}^\infty b_s \varepsilon^s ,
\\
  \frac{ n! }{ \Gamma ( n + 1 - \varepsilon ) }
&\sim
  \sum_{t=0}^\infty c_{t,n} \varepsilon^s .
\end{align}
Expressions for $a_r$, $b_s$ are readily found,
\begin{align}\label{n_toro:abcexpressions}
\nonumber
  a_r
&=
  \frac{ ( - \ln y_0 )^r }{ r! } ,
\\
  b_s
&=
  \left\{
    \begin{array}{cc}
      {\displaystyle
        \frac{
          2 ( 2^{s-1} - 1 ) \pi^s \lvert \mathrm{B}_s \rvert
        }{
          s!
        } ,
      }
      & s\text{ even}, \\
      &\\
      0 ,
      & s\text{ odd}, \\
    \end{array}
  \right.
\end{align}
where $\mathrm{B}_s$ are Bernoulli numbers.  As for $c_{t,n}$, it can
be written as a sum over Young tableaux of order $t$,
\begin{equation}\label{n_toro:cyoungtableaux}
  c_{t,n}
=
  \sum_{\mathrm{Y.T.}(t)}
  ( -1 )^{ t - \sum_{j=1}^t \nu_j }
  \frac{
        \prod_{i=1}^t \left[ \psi^{(i-1)} (n+1) \right]^{\nu_i}
      }{
        \prod_{\ell=1}^t \ell!^{\nu_\ell} \, \nu_\ell!
       },
\end{equation}
where $\nu_j$, $j=1, \, \ldots, \, t,$ are the numbers of rows with
$j$ elements, such that $\sum_{j=1}^t j \nu_j = t$, and $\psi$ is the
digamma function.

As an example, for the $k=1$ sector in the \cpuno{} model we need
\begin{equation}\label{n_toro:tunoene}
  T_{1,n}
=
  \psi (n+1) - \ln y_0,
\end{equation}
from which equation (\ref{n_toro:sumaresiduosnkdos}) obtains.  For the
$k=2$ sector in the same model, expression
(\ref{n_toro:cyoungtableaux}) is already too cumbersome to compute
(\ref{n_toro:tresiduo}) explicitly:
\begin{align}\label{n_toro:ttresene}
\nonumber
  T_{3,n}
&=
 - \, \frac{1}{6} ( \ln y_0 )^3
 - \, \frac{ \pi^2 }{6} \ln y_0
 + \left[
         \frac{1}{2} ( \ln y_0 )^2 + \frac{ \pi^2 }{6}
   \right]
   \psi (n+1)
\\
\nonumber
&
\qquad
 - \, ( \ln y_0 )
   \left(
         \frac{1}{2} \psi(n+1)^2 - \psi'(n+1)
   \right)
\\
&
\qquad
 + \frac{1}{6} \, \psi(n+1)^3
 - \, \frac{1}{2} \, \psi(n+1) \psi'(n+1)
 + \frac{1}{6} \, \psi''(n+1) .
\end{align}

However, it is not necessary to perform the summation in
(\ref{n_toro:sumandoespero}) to realise that the pole structure has a
natural interpretation in terms of instantons.  From the original
classical action
\begin{equation}\label{n_toro:accionconamuyalfaeuclidea}
  S[\Psi, \, \Psi^\dagger, \, A_\mu, \, \alpha]
=
  \frac{N}{2g_0^2}
  \int_{\mathrm{T}^2} \mathrm{d}^2 x \,
    \lvert \mathrm{D}_\mu \Psi \rvert^2
 +
  \frac{N}{2g_0^2}
  \int_{\mathrm{T}^2} \mathrm{d}^2 x \,
    \alpha (x)
    \left( \Psi^\dagger \Psi - 1 \right) ,
\end{equation}
the classical equations of motion
\begin{equation}\label{n_toro:ecmovimientoclasicas}
  \left(
        - \mathrm{D}_\mu^2 + \alpha
  \right)
  \Psi
=
  0 ,
\quad
  \alpha
=
  \Psi^\dagger 
  \mathrm{D}_\mu^2 \Psi
\end{equation}
ensure that, for classical solutions, the value of the action is given
by the integral of $- \alpha$:
\begin{equation}\label{n_toro:valordelaaccionclasicaensoluciones}
  S_{\mathrm{cl}}
=
  \frac{N}{2g_0^2}
  \int_{\mathrm{T}^2} \mathrm{d}^2 x \,
    \lvert \mathrm{D}_\mu \Psi \rvert^2
=
  \frac{N}{2g_0^2}
  \int_{\mathrm{T}^2} \mathrm{d}^2 x \,
  ( - \alpha ) .
\end{equation}
The $n$th pole of the integrand of $Z_k$ corresponds to a value of
$\alpha = m^2$ such that
\begin{equation}\label{n_toro:posicionpoloscomoinstantones}
  \frac{ m^2 L^2 }{ 4 \pi \lvert k \rvert }
=
  - \left( n + \frac{1}{2} \right)
\ \Longrightarrow \ {}
  S_{\mathrm{cl}}
=
  \frac{N}{2g_0^2} \,
  m^2 L^2
=
  ( 1 + 2 n ) \,
  \frac{N}{2g_0^2} \,
  2 \pi \lvert k \rvert ,
\end{equation}
i.e., it exactly matches the classical action of a
\textsl{multiinstanton} configuration composed of an instanton and $n$
instanton-antiinstanton pairs (figure \ref{n_toro:fig:cerosypolos}).
This is in contrast with the case of the sphere, where the structure
of poles of the integrand of $Z_k$ does not correspond to charge $k$
multiinstanton configurations.

Notice that, although there are no unit charge instantons on the torus
\cite{Braam:1989qk}, there is a nontrivial contribution of the $k =
\pm 1$ sectors to the total partition function.  This is so because
the dominant configurations in the partition function are not the
smooth classical solutions, but rather the rough configurations in the
neighbourhood of these.  The classical solution fails to exist for
unit charge, but this is a classical accident, which does not affect
the quantum dynamics of the system because of the nonzero contribution
of the neighbouring configurations.

\subsubsection{\label{subsubsec:torus:saddle}%
  The saddle point method}

Now let us consider the extrema of $\tilde{S} ( y )$.  These are the
zeros of its derivative,
\begin{equation}\label{n_toro:derivadadestilde}
  \frac{
        \mathrm{d} \tilde{S} (y)
      }{
        \mathrm{d} y
       }
=
  \ln y_0
 -
  \psi \left( y + 1/2 \right) ,
\end{equation}
and constitute a sequence $y = s_n$, $n = 0, \, 1, \, \ldots$ Saddle
points and poles alternate, $s_0 > p_0 > s_1 > p_1 > \cdots$, as seen
in figures \ref{n_toro:fig:sdeytoro} and \ref{n_toro:fig:cerosypolos}.
The saddle points approach the poles for large $n$, $\lim_{n
  \rightarrow \infty} s_n / p_n = 1$.

If the dominant saddle point $s_0$ lies in the region $y \gg 1$, we
can find its location as an expansion in powers of $y_0^{-1}$ starting
from (\ref{n_toro:exponenteyengrany}):
\begin{equation}\label{n_toro:desarrollosceroenygrande}
  s_0
=
  y_0
 +
  \frac{ 1 }{ 24 \, y_0 }
 - \,
  \frac{ 12097 }{ 576 \, y_0^3 }
 +
  \mathcal{O} \left( y_0^{-5} \right) .
\end{equation}
Equivalently, the infinite volume value $m_{\mathrm{T}}^2$ of the
saddle point receives finite volume corrections,
\begin{equation}\label{n_toro:correccionessillavolfinito}
  m_{s_0}^2
=
  m_{\mathrm{T}}^2
  \left(
        1
       +
        \frac{ 2 \pi^2 \lvert k \rvert }{ 3 m_{\mathrm{T}}^4 L^4 }
       +
        \mathcal{O} \left( y_0^{-6} \right)
  \right) .
\end{equation}

We evaluate the partition function in sector $k$, up to quadratic
order,
\begin{equation}\label{n_toro:accionpuntosillacuadratico}
  Z_k^{(s_0)}
\approx
  \frac{ 4 \pi \lvert k \rvert }{ L^2 } \,
  \sqrt{ \frac{ 2 \pi }{ N \lvert k \rvert } } \,
  \mathrm{e}^{ - N \lvert k \rvert \, \tilde{S} (s_0) } \,
  \left(
        \frac{ \mathrm{d}^2 \tilde{S} }{ \mathrm{d} y^2 }
  \right)_{ \!\! s_0 }^{ \!\! -1/2 } ,
\end{equation}
from which
\begin{align}\label{n_toro:fparticionksaddlegrany}
\nonumber
  Z_k^{(s_0)}
&=
  \frac{ 4 \pi \lvert k \rvert }{ L^2 } \,
  \sqrt{ \frac{ 2 \pi y_0 }{ N \lvert k \rvert } } \,
  \exp \left\{
         - N \lvert k \rvert \, y_0
         - \, \frac{ N \lvert k \rvert }{ 24 \, y_0 }
         + \frac{ 1 }{ 16 y_0^2 }
         + \frac{ 29 \, N \lvert k \rvert }{ 5760 y_0^3 }
         + \mathcal{O} \left( y_0^{-5} \right)
       \right\}
\\
&=
  \frac{ 4 \pi m_{\mathrm{T}} }{ \sqrt{2N} L } \,
  \exp \left\{
        - N \frac{ m_{\mathrm{T}}^2 L^2 }{ 4 \pi }
        - \, \frac{ \pi }{ 6 } \,
             \frac{ N k^2 }{ m_{\mathrm{T}}^2 L^2 }
        + \left(
            \frac{ \pi \lvert k \rvert }{ m_{\mathrm{T}}^2 L^2 }
          \right)^2
        + \mathcal{O} \left( y_0^{-3} \right)
       \right\} .
\end{align}

The partition function in the trivial topological sector can be
computed in the large $m^2 L^2$ limit by substituting an integral for
the sum in the definition of the $\zeta$ function,
\begin{equation}\label{n_toro:fparticionsectorcero}
  Z_0
=
  \int \mathrm{d} m^2 \,
  \exp \left\{
         N \,
         \frac{ m^2 L^2 }{ 4 \pi } \,
         \left(
           \ln \frac{ m^2 }{ m_{\mathrm{T}}^2 } - 1
         \right)
       \right\}
\; ,
\end{equation}
and coincides with the $B \rightarrow 0$ limit of the result for
$Z_k$.  There are no poles in this sector, which is compatible with
the absence of instantons of zero topological charge, and its only
saddle point yields an approximation
\begin{equation}\label{n_toro:fparticionsectorceropuntosilla}
  Z_0^{(s_0)}
\approx
  \frac{ 4 \pi m_{\mathrm{T}} }{ \sqrt{2N} L } \,
  \exp \left\{
             - N \frac{ m_{\mathrm{T}}^2 L^2 }{ 4 \pi }
       \right\} ,
\end{equation}
agreeing with the result (\ref{n_toro:fparticionksaddlegrany}) for the
nontrivial sectors.

The full partition function in terms of the vacuum angle $\theta$, in
the saddle point approximation, is the sum
\begin{align}\label{n_toro:fparticiontotalsaddlegrany}
\nonumber
  Z^{(s_0)} ( \theta )
&=
  \sum_{ k \in \mathbb{Z} }
  Z_k^{(s_0)} \mathrm{e}^{ - i k \theta}
\\
\nonumber
&\approx
  \frac{ 4 \pi m_{\mathrm{T}} }{ \sqrt{2N} L } \,
  \exp \left\{
             - N \frac{ m_{\mathrm{T}}^2 L^2 }{ 4 \pi }
       \right\} \,
  \sum_{ k \in \mathbb{Z} }
  \exp \left\{
         - \, \frac{ \pi }{ 6 } \,
           \frac{ N k^2 }{ m_{\mathrm{T}}^2 L^2 }
         - i k \theta
       \right\}
\\
&=
  \frac{ 4 \pi m_{\mathrm{T}} }{ \sqrt{2N} L } \,
  \exp \left\{
             - N \frac{ m_{\mathrm{T}}^2 L^2 }{ 4 \pi }
       \right\} \,
  \vartheta_3
    \left(
      \frac{ \theta }{ 2 \pi }
    \right\vert \left.
      \frac{ i N }{ 6 m_{\mathrm{T}}^2 L^2 }
    \right) ,
\end{align}

Still in the region $y \gg 1$, i.e. $m^2 L^2 \gg 4 \pi \lvert k
\rvert$, let us analyse this partition function in two different
regimes, depending on the relation between the number $N$ and
$m_{\mathrm{T}}^2 L^2$.

For $N \gg m_{\mathrm{T}}^2 L^2$, the contribution of high topological
sectors (large $\lvert k \rvert$) to
(\ref{n_toro:fparticiontotalsaddlegrany}) can be neglected.  Keeping
just sectors $k = 0, \, \pm 1$, we obtain
\begin{equation}\label{n_toro:fparticiontruncagrann}
  Z^{(s_0)} ( \theta )
\approx
  \frac{ 4 \pi m_{\mathrm{T}} }{ \sqrt{2N} L } \,
  \exp \left\{
             - N \frac{ m_{\mathrm{T}}^2 L^2 }{ 4 \pi }
       \right\} \,
  \left(
         1
        +
         2 \,
         \exp \left\{
               - \, \frac{ \pi }{ 6 } \,
               \frac{ N }{ m_{\mathrm{T}}^2 L^2 }
              \right\}
         \cos \theta
  \right) ,
\end{equation}
giving rise to a vacuum energy density
\begin{align}\label{n_toro:densenergiavaciotruncagrann}
\nonumber
  \mathcal{E}_0^{(s_0)} ( \theta )
 -
  \mathcal{E}_0^{(s_0)} ( 0 )
&=
 - \, \frac{ 1 }{ L^2 } \,
  \ln \frac{ Z^{(s_0)} ( \theta ) }{ Z^{(s_0)} ( 0 ) }
\\
&\approx
  \frac{ 4 }{ L^2 } \,
  \exp \left\{
         - \, \frac{ \pi }{ 6 } \,
           \frac{ N }{ m_{\mathrm{T}}^2 L^2 }
       \right\}
  ( 1 - \cos \theta ) ,
\end{align}
i.e., the typical $\mathcal{E}_0 ( \theta )$ dependence of a dilute
instanton gas.  This is a $2 \pi$-periodic function, smooth for all
values of $\theta$ including $\theta = \pm \pi$.  Hence, there is no
first order phase transition.

This regime would be compatible with a \textsl{definition} of the
$1/N$ expansion in which both the large $N$ limit and the $1/N$
corrections are studied in finite volume.

However, if we move to the region $m_{\mathrm{T}}^2 L^2 \gg N$, it
proves convenient to use the Poisson formula in
(\ref{n_toro:fparticiontotalsaddlegrany}) to get
\begin{align}\label{n_toro:fparticiontotalpoisson}
\nonumber
  Z^{(s_0)} ( \theta )
&=
  \frac{ 4 \sqrt{3} \, \pi m_{\mathrm{T}}^2 }{ N } \,
  \exp \left\{
             - N \frac{ m_{\mathrm{T}}^2 L^2 }{ 4 \pi }
       \right\} \,
  \vartheta
    \left[
      \begin{array}{c}
      \frac{ \theta }{ 2 \pi } \\
      0
      \end{array}
    \right]
    \left(
      0
    \left\vert
      \frac{ i 6 m_{\mathrm{T}}^2 L^2 }{ N }
    \right. \right) ,
\\
&=
  \frac{ 4 \sqrt{3} \, \pi m_{\mathrm{T}}^2 }{ N } \,
  \exp \left\{
             - N \frac{ m_{\mathrm{T}}^2 L^2 }{ 4 \pi }
       \right\} \,
  \sum_{ q \in \mathbb{Z} }
  \exp \left\{
         - \, \frac{ 6 \pi m_{\mathrm{T}}^2 L^2 }{ N } 
         \left( q + \frac{\theta}{2\pi} \right)^2
       \right\} ,
\end{align}
where the $\vartheta$ function with characteristics is given by
\begin{equation}\label{n_toro:varthetachars}
  \vartheta
    \left[
      \begin{array}{c}
      a \\ b
      \end{array}
    \right]
    ( z \vert \tau )
=
  \sum_{ n \in \mathbb{Z} }
  \exp \left\{
           i \pi \tau ( n + a )^2 
         + i 2 \pi ( n + a ) ( z + b )
       \right\} .
\end{equation}
Now, it is the dual sum in $q$ that is dominated by the low $\lvert q
\rvert$ terms. Defining again $\tilde{\theta}$ as the angle in $[ -
  \pi, \, + \pi ]$ differing from $\theta$ by an integer,
\begin{equation}\label{n_toro:fparticiontotalpoissonaprox}
  Z^{(s_0)} ( \theta )
\approx
  \frac{ 4 \sqrt{3} \, \pi m_{\mathrm{T}}^2 }{ N } \,
  \exp \left\{
         - N \frac{ m_{\mathrm{T}}^2 L^2 }{ 4 \pi }
         - \, \frac{ 3 m_{\mathrm{T}}^2 L^2 }{ 2 \pi N } \,
           {\tilde{\theta}}^2
       \right\} .
\end{equation}
This reproduces the traditional large $N$ picture, where instanton
effects are suppressed, and the vacuum energy density depends
quadratically on $\theta$ within the interval $[ - \pi, \, + \pi ]$:
\begin{equation}\label{n_toro:densenergiavaciotruncagranv}
  \mathcal{E}_0^{(s_0)} ( \theta )
 -
  \mathcal{E}_0^{(s_0)} ( 0 )
\approx
  \frac{ 3 m_{\mathrm{T}}^2 }{ 2 \pi N } \, {\tilde{\theta}}^2 .
\end{equation}

Periodicity in $\theta$ is guaranteed because $\mathcal{E}_0 ( \theta
)$ depends on the periodic variable $\tilde{\theta}$, but this
function is not smooth at odd multiples of $\pi$, where
$\tilde{\theta}$ is doubly defined, levels cross and a first order
phase transition occurs.

This regime is compatible with a definition of the $1/N$ expansion in
which the thermodynamic limit is performed first, and $N$ is taken to
infinity afterwards.

These results agree with the analysis of \cpn{} models on the sphere.
Since the physical pictures pertaining to the regimes $N \gg
m_{\mathrm{T}}^2 L^2$ and $m_{\mathrm{T}}^2 L^2 \gg N$ are different,
the limits $N \rightarrow \infty$ and $V \rightarrow \infty$ do not
commute, and the orders in which these limits are taken determine
different theories. This behaviour points towards the existence of a
phase transition in the $N\to \infty$ theory.

%
\section{\label{sec:discussion}%
  Discussion}

After the careful analysis of the large $N$ method on the sphere and
the torus, we conclude that the apparent incompatibility between
instanton physics and the $1/N$ expansion has its cause in the
formulation of the latter in infinite volume and is a subtle effect of
the noncommutativity of the large $N$ and thermodynamic limits.

To clarify this, consider the essential dependence of the vacuum
energy density on the angle $\theta$, the volume $V$, and the number
of colours $N$, at fixed saddle point mass $m_{\mathrm{T}}$
\begin{equation}\label{discussion:essential}
  \mathcal{E}_0 ( \theta )
=
 - \,
  \frac{ 1 }{ V }
  \ln
  \frac{
    \vartheta_3
    \left(
      \frac{ \theta }{ 2 \pi }
    \left\vert
      \frac{ i N }{ 6 m_{\mathrm{T}}^2 V }
    \right.\right)
  }{
    \vartheta_3
    \left(
      0
    \left\vert
      \frac{ i N }{ 6 m_{\mathrm{T}}^2 V }
    \right.\right)
  } ,
\end{equation}
which is valid in the cases of the sphere and the torus provided we
\textsl{define} the free energy as a function of $\theta$ by
subtraction of the contribution at $\theta = 0$ for each $N$.

The various limits of (\ref{discussion:essential}) are best discussed
in terms of a dimensionless variable $x \equiv N / ( 6
m_{\mathrm{T}}^2 V )$ and the function
\begin{equation}\label{discussion:essentialdimless}
  \frac{ N }{ 6 m_{\mathrm{T}}^2 }
  \mathcal{E}_0 ( \theta )
\equiv
  f ( \theta, \, x )
=
 - \,
  x
  \ln
  \frac{
    \vartheta_3
    \left(
      \frac{ \theta }{ 2 \pi }
    \left\vert
      i x
    \right.\right)
  }{
    \vartheta_3
    \left(
      0
    \left\vert
      i x
    \right.\right)
  }
=
 - \,
  x
  \ln
  \frac{
    \vartheta
    \left[
      \begin{array}{c}
        \theta / (2 \pi) \\ 0
      \end{array}
    \right]
    \left(
      0
    \left\vert
      \frac{ i }{ x }
    \right.\right)
  }{
    \vartheta_3
    \left(
      0
    \left\vert
      \frac{ i }{ x }
    \right.\right)
  } .
\end{equation}
The last equation is obtained by applying the Poisson resummation
formula, that is, the modular transformation of the theta functions.
For $x$ strictly positive, $f$ is a well defined real analytic
function of $\theta \in \mathbb{R}$.  The nonanalyticities for complex
$\theta$ are branch cuts located at the zeros of the Jacobi
$\vartheta_3$ function, that is, for $\theta = ( 1 + i x ) \pi$ and
its translations by integer multiples of $2 \pi$ and of $i 2 \pi x$.
These zeros never occur for real values of $\theta$.

The two limiting regimes we have been discussing are given in terms of
the dimensionless quantities as:
\begin{itemize}
\item %
  Semiclassical: send $x$ to infinity (meaning $N \gg m_{\mathrm{T}}^2
  V$).  In this case
  \begin{equation}\label{discussion:dimlesssemiclassicalapproxtheta}
    \vartheta_3
    \left( \left.
      \frac{ \theta }{ 2 \pi }
    \right\vert
      i x
    \right)
  =
    1
   +
    2 \, \mathrm{e}^{ - \pi x } \cos \theta
   +
    \mathcal{O} ( \mathrm{e}^{ - 2 \pi x } ) ,
  \end{equation}
  and we recover the nonperturbative result
  \begin{equation}\label{discussion:essentialdimlesssemiclassical}
    f ( \theta, \, x )
  =
    x \,
    \mathrm{e}^{ - \pi x }
    ( 1 - \cos \theta )
   +
    \mathcal{O} ( \mathrm{e}^{ - 2 \pi x } ) ,
  \end{equation}
  equivalent to
  \begin{equation}\label{discussion:essentialdimlesssemiclassicalphys}
    \mathcal{E}_0 ( \theta )
  =
    \frac{ 1 }{ V } \,
    \exp \left\{ - \, \frac{ \pi N }{ 6 m_{\mathrm{T}}^2 V } \right\} \,
    ( 1 - \cos \theta )
   +
    \mathcal{O}
    \left(
      \frac{ m_{\mathrm{T}}^2 }{ N } \,
      \exp \left\{ - \, \frac{ \pi N }{ 3 m_{\mathrm{T}}^2 V } \right\}
    \right) ,
  \end{equation}
  equivalent to (\ref{tworegimes:semiclassical}) (and vanishing as $x
  \rightarrow \infty$ together with all of its derivatives).
\item %
  Traditional large $N$: send $x$ to zero (meaning $N \ll
  m_{\mathrm{T}}^2 V$).  This is a problematic limit for the modular
  parameter $\tau = i x$, which leaves the upper half plane.  We
  consider separately the regions $\theta \in ( -\pi, \, \pi )$ and
  $\theta = \pi$ (the rest of the function is obtained by
  periodicity):
  \begin{align}\label{discussion:essentialdimlesslargen}
  \nonumber
    f \big( \theta \in ( -\pi, \, \pi ), \, x \big)
  &=
    \frac{ \theta^2 }{ 4 \pi }
   +
    \mathcal{O}
    \left(
      \exp \left\{
            - \, \frac{ \pi ( 1 - \theta / \pi )^2 }{ x }
           \right\}
    \right) ,
  \\
    f ( \theta = \pi, \, x )
  &=
    \frac{ \pi^2 }{ 4 \pi }
   -
    x \ln 2
   +
    \mathcal{O} ( \mathrm{e}^{ - \pi / x } ) ,
  \end{align}
  reproducing (\ref{tworegimes:largen}), including the first order
  cusp at $\theta = \pi$:
  \begin{align}\label{discussion:essentialdimlesslargenphys}
  \nonumber
    \mathcal{E}_0 \big( \theta \in ( -\pi, \, \pi ) \big)
  &=
    \frac{ 3 m_{\mathrm{T}}^2 }{ 2 \pi N } \,
    \theta^2
   +
    \mathcal{O}
    \left(
      \frac{ m_{\mathrm{T}}^2 }{ N } \,
      \exp \left\{
             - \, \frac{ 6 \pi m_{\mathrm{T}}^2 V \,
               ( 1 - \theta / \pi )^2 }{ N }
            \right\}
    \right) ,
  \\
    \mathcal{E}_0 ( \theta = \pi )
  &=
    \frac{ 3 m_{\mathrm{T}}^2 }{ 2 \pi N } \,
    \pi^2
   -
    \frac{ \ln 2 }{ V }
   +
    \mathcal{O}
    \left(
      \frac{ m_{\mathrm{T}}^2 }{ N } \,
      \exp \left\{
             - \, \frac{ 6 \pi m_{\mathrm{T}}^2 V }{ N }
           \right\}
    \right) ,
  \end{align}

\end{itemize}

The r\^ole of $i x$ as a modular parameter suggests an analogy with
finite temperature models, where the number of colours corresponds to
the inverse temperature $\beta$.  In this sense, we expect that the
physics at $1/N = 0$ corresponds to zero temperature phenomena.  The
fact that the thermodynamic and large $N$ limits do not commute
(reflected in the behaviour of $x$) would suggest the presence of a
phase transition exactly at zero temperature, i.e., $1/N = 0$.

Returning to physical quantities, we have shown that finite volume
effects in the $\theta$ dependence of the vacuum energy density for
\cpn{} models on S$^2$ and T$^2$, when all topological sectors are
taken into account, give rise to two asymptotic regimes, one dominated
by instanton effects (when $N \gg m_{\mathrm{T}}^2 V$) and the other
by the conventional large $N$ picture (when $N \ll m_{\mathrm{T}}^2
V$). These are smoothly connected by an interpolating region.

It should be realised that the basic hypotheses of the method of large
$N$ do not hold when $N \ll m_{\mathrm{T}}^2 V$, for which precisely
the traditional large $N$ results obtain.  The saddle point technique
needs $N$ to be the largest dimensionless parameter of the theory, in
particular larger than $m_{\mathrm{T}}^2 V$.  Two very different
theories are defined by interchanging the noncommuting limits $N
\rightarrow \infty$ and $V \rightarrow \infty$.  In principle, the
only procedure consistent with the saddle point method is taking the
large $N$ limit first, and then going over to the thermodynamic limit.

However, for small values of $\theta$, where the large $N$
approximation is expected to hold \cite{Coleman:1976uz}, both
procedures seem to make sense.  Lattice measurements of the $\theta =
0$ topological susceptibility agree with the traditional large $N$
picture, corresponding to performing first the $V \rightarrow \infty$
limit, and then taking $N$ to infinity.  However, these simulations
have been only carried out for values of the parameters such that
$m_{\mathrm{T}}^2 V \gtrsim N$ (see Table \ref{table:gap} and
references \cite{burk,acep,Campostrini:1993wy}), agreeing with our
analysis in the region not validated by the saddle point method.  It
would be interesting to rerun these simulations for smaller volumes,
still close to the thermodynamic limit with a stable mass gap, but
where an instanton-dominated $\theta$ dependence of the vacuum energy
density could emerge and eventually take over.  Notice that the
singularity of the topological susceptibility pointed out by lattice
simulations in the \cpuno{} model is also found in the semiclassical
scenario.  The existence of such a singularity can also be understood
by the presence of a family of Lee-Yang zeros of the analytic
continuation of the partition function in the complex $\theta$-plane,
converging to $\theta=0$ in the thermodynamic limit \cite{aa2009}.

\begin{table}[h]
  \renewcommand{\arraystretch}{1.1}
  \begin{center}
  \begin{tabular}{|l|r|r|r|r|}
    \hline  
    Reference  &  $m$  &  $L$  &  $m^2 L^2$  &  $N$ \\
    \hline\hline
      \cite{Hasenfratz} \,\         Blatter {\it $\!$et $\!$al.}
       & 0.165 &  40 &  43.6 &  2 \\
      \cite{Ahmad} \,\              Ahmad {\it $\!$et $\!$al.}
       & 0.179 &  50 &  80.1 &  2 \\
      \cite{lian} \,\               Lian-Thacker 
       & 0.111 & 100 & 123.2 &  2 \\
      \cite{kh} \,\                 Keith-Hynes-Thacker
       & 0.179 &  50 &  80.1 &  2 \\
      \cite{lian} \,\               Lian-Thacker
       & 0.084 & 100 &  70.6 &  3 \\
      \cite{Campostrini:1992it} \,\ Campostrini {\it $\!$et $\!$al.}
       & 0.066 & 120 &  62.7 &  4 \\
      \cite{BurkJap} \,\            Burkhalter {\it $\!$et $\!$al.}
       & 0.131 &  32 &  17.6 &  4 \\
      \cite{lian} \,\               Lian-Thacker
       & 0.088 & 100 &  77.4 &  4 \\
      \cite{kh} \,\                 Keith-Hynes-Thacker
       & 0.180 &  50 &  81.0 &  4 \\ 
      \cite{Ahmad} \,\              Ahmad {\it $\!$et $\!$al.}
       & 0.186 &  50 &  86.5 &  6 \\ 
      \cite{lian} \,\               Lian-Thacker
       & 0.085 & 100 &  72.2 &  6 \\
      \cite{Campostrini:1992it} \,\ Campostrini  {\it $\!$et $\!$al.} 
       & 0.196 &  72 & 199.1 & 10 \\
      \cite{Debbio} \,\             Del Debbio  {\it $\!$et $\!$al.}
       & 0.192 &  60 & 132.7 & 10 \\
      \cite{Ahmad} \,\              Ahmad  {\it $\!$et $\!$al.}
       & 0.212 &  50 & 112.4 & 10 \\
      \cite{lian} \,\               Lian-Thacker
       & 0.058 & 100 &  33.6 & 10\\
      \cite{kh} \,\                 Keith-Hynes-Thacker
       & 0.212 &  50 & 112.4 & 10 \\
      \cite{Debbio} \,\             Del Debbio  {\it $\!$et $\!$al.}
       & 0.397 &  42 & 278.0 & 15 \\
      \cite{Debbio} \,\             Del Debbio  {\it $\!$et $\!$al.}
       & 0.418 &  30 & 157.3 & 21 \\
      \cite{vicari} \,\             Vicari
       & 0.287 &  60 & 296.5 & 21 \\
      \cite{Debbio} \,\             Del Debbio  {\it $\!$et $\!$al.}
       & 0.305 &  56 & 291,8 & 21 \\
      \cite{vicari} \,\             Vicari
       & 0.411 &  42 & 298.0 & 41 \\
   \hline
\end{tabular}
\end{center}
\caption{\label{table:gap} Numerical values of the mass gap $m$ for
  different \cpn{} models at different volumes $m^2 L^2$.  The models
  are in the thermodynamic regime $m^2 L^2 > N$ in all cases. The
  r\^ole of instantons is only manifest in the cases \cpuno{} and
  \cpdos{} \cite{lian}}
\end{table}

As regards the neighbourhood of $\theta = \pi$, the nodal analysis of
\cite{Asorey:1998cz} appears incompatible with a first order phase
transition at $\theta = \pi$.  This is the behaviour of the system for
lower values of $N$ for any volume, i.e. \cpuno{} and \cpdos{} models.
For larger values of $N$, for instance $N>4$, this behaviour is only
observed for small volumes, i.e. volumes which verify $m^2 V \ll N$.
For larger volumes the effect is swept off by the infrared
fluctuations and the system undergoes a phase transition at
$\theta=\pi$ with spontaneous CP symmetry breaking.

The behaviour for intermediate values of $\theta$ has so far proved
elusive to numerical techniques, due to the inaccuracies inherent to
lattice simulations in this region. Let us however remark that a novel
technique \cite{Azcoiti:2002vk} based on analytically continuing the
$\theta$ dependence to imaginary values has been introduced to
overcome this problem.  This technique has been applied to the
$\mathbb{C} \mathrm{P}^9$ model \cite{Azcoiti:2003pendiente}, with
conclusions agreeing with the usual large $N$ expansion.  The
consistency of this technique for any value of $N$ is supported by the
absence of singularities in the analytic extension of the partition
function to the whole $\theta$-plane \cite{aa2009}.

Summing up, the large $N$ method is compatible with instanton effects.
Besides, the results on $\theta$ dependence obtained with this tool
agree with the known behaviour at $\theta = 0$, i.e., the vacuum
energy density is differentiable there and the Vafa-Witten theorem
holds.  It is also compatible with the numerical determination of the
topological susceptibility at $\theta = 0$.  The analysis of the poles
of the partition function on the torus supports the method of Jevicki,
whereby instantonic effects appear in the large $N$ limit in the form
of resonances.  First order phase transitions with spontaneous parity
breaking at $\theta = \pi$ appear in the formulation of the models
directly at infinite volume, and we have exposed the analytic roots of
this fact.  The large $N$ method is thus in agreement with all exact
results on $\theta$ dependence, and provides a valuable bridge between
the angles $\theta = 0$ and $\theta = \pi$.

Let us remark that the behaviour of the theory in finite volume plays
a fundamental r\^ole in condensed matter settings, where sigma models
can be used as effective theories for the quantum Hall effect
\cite{Levine:1983vg, Pruisken:1984ni, Khmelnitskii:1983qhef,
  Levine:1984yg}.  In this context, the Hall conductivity is
identified with the coupling of the topological term, and the
stability of Hall plateaux is linked to the renormalisation group
running of the couplings (including $\theta$).  The large $N$ limit of
\cpn{} models was studied in connection with this phenomenon in a
series of papers (see, e.g., \cite{Pruisken:2000my,
  pruisken2004burmistrov, pruisken2005shankar, burmistrov2007,
  pruisken2008}), in which the different regimes we have discussed
were also identified; in this case the traditional large $N$ limit at
infinite volume is blind to edge effects, which of course are crucial
for the physics of the Hall effect. In particular, edge currents are a
finite size effect and this suggests that the $m^2 V \ll N$ regime of
\cpn{} sigma models is the relevant regime for their description.
  
Finally, we remark that the difference between the two regimes is due
to the asymptotic behaviour of the effective action in the different
topological sectors.  According to equations
(\ref{n_toro:exponenteyengranyphysical}) and
(\ref{n_toro:exponenteyenpequenyoyphysical}), for small values of the
topological charge, $\lvert k \rvert < m^2 V$, the effective action is
quadratic in $k$, whereas for large topological charges its leading
term is linear in $\lvert k \rvert$ \cite{aa2009}.  The two regimes
also differ at finite temperature.  Since the spacetime volume is $V =
LT$, the change of asymptotic dependence of the effective action on
the topological charge can be associated with a finite temperature
crossover from the low temperature regime $\beta = 1/T > m^2L/|q|$ to
the high temperature regime $\beta < m^2 L/|q|$ and cannot be related
to any phase transition \cite{affl}. One might expect a similar
phenomenon in QCD, although in that case there is a finite temperature
phase transition \cite{kpt}.

%
\subsection*{Acknowledgements}
We thank D.~Garc\'{\i}a-Alvarez for discussions on the early stages of
this paper.  M.~Aguado thanks Departamento de F\'{\i}sica Te\'orica of
the University of Zaragoza for hospitality in visits where this
project was developed.  M.~Asorey was partially supported by the
Spanish CICYT grant FPA2009-09638 and DGIID-DGA (grant 2009-E24/2).

%
%

%
\end{document}